At extreme strain rates, pure metals thermally harden while alloys thermally soften

**Authors:** Ian Dowding[1], Christopher A. Schuh[1,2*]

**Affiliations:**

[1]Department of Materials Science and Engineering, Massachusetts Institute of Technology; Cambridge, MA 02139, USA.

[2]Department of Materials Science and Engineering, Northwestern University; Evanston, IL 60208, USA.

*Corresponding author. Email: schuh@northwestern.edu

**Abstract**

When materials are deformed at extreme strain rates, $>10^6$ s$^{-1}$, a counterintuitive mechanical response is seen where the strength and hardness of pure metals increases with increasing temperature. This anti-thermal hardening is due to defects in the material becoming pinned by phonons in the crystal lattice. However, here, using optically-driven microballistic impact testing to measure dynamic strength and hardness, we show that when the composition is systematically varied away from high purity, the mechanical response of metals transitions from ballistic transport of dislocations back to thermally activated pinning of dislocations, even at the highest strain rates. This boundary from "hotter-is-stronger" to "hotter-is-softer" is observed and mapped for nickel, titanium and gold. The ability to tune between deformation mechanisms with very different temperature dependencies speaks to new directions for alloy design in extreme conditions.

**Introduction**

Most materials thermally soften with increasing temperature, because temperature allows for easier slippage of bonds, and the motion of strain-carrying defects like dislocations. Such thermal softening has been studied extensively and quantitively over 10 orders of magnitude of strain rates, and it prevails in materials tested up to ~$10^5$ s$^{-1}$. This is already a very high strain rate, relevant to car crashes (*1*), hard particle erosion (*2*), ballistic impacts (*3*), and manufacturing processes (*4*), and lying at the upper range of common high–rate mechanical tests including gas gun impacts (*5*), flyer plate impacts (*6*), and Kolsky bar impacts (*7*). To reach higher strain rate deformations (above $10^5$ s$^{-1}$), high pressure shock experiments (laser shock impacts, plate impacts) are often used, which generate strong shocks in the material, and in turn lead to shock conflation with the measured properties. Lower pressure experiments, such as Richtmyer-Meshkov instability (RMI) experiments (*8*), can produce extreme deformations at rates of ~$10^7$ s$^{-1}$, but infer properties through hydrodynamic instabilities and do not provide directly measured quantities. Recent advances in microparticle impact testing, however, have made it possible to micromechanically test materials using dynamic indentation tests at strain rates above $10^6$ s$^{-1}$ and at low pressures of just a few GPa, well below where strong shock effects and hydrodynamic instabilities become significant (*9–14*).

Our recent work (*9*) experimentally showed that at these extreme strain rates above $10^5$ s$^{-1}$ and at low pressures, pure Cu, Au, and Ti increase strength with increasing temperature; they do not thermally soften, but rather thermally harden. This effect is due to kinetic limitations on defect motion and a change in deformation mechanism at extreme strain rates. The thermally activated transport of dislocations at lower rates gives way to ballistic transport of dislocations limited by dislocation-phonon interactions (*15*). This transition leads to "anti-thermal" behavior where there is an increase in strength with increasing temperature, because temperature inflates the number of pinning points created by phonons. While this mechanism is dominant in high purity metals, it is always in competition with other deformation mechanisms, and it is not yet clear what its range of dominance will be in the face of, e.g., added alloying elements, which introduce new strengthening mechanisms not present in pure metals. At low strain rates, alloying leads to additional dislocation pinning mechanisms; at the highest strain rates where phonon drag is relevant, however, such alloying effects are not well quantified. Here, using microparticle impact testing, we systematically explore several orders of magnitude of solid-solution additions into pure metals. We show that phonon drag crosses from dominant to secondary as a rate-limiting mechanism as the solute level rises: Only high purity metals strengthen with increasing temperature, and there is a clear crossover to thermally activated flow at lower purity.

**Extreme Strain Rate Testing on Nickel Alloys**

Impact experiments were carried out using high velocity microparticle impacts with the laser-induced particle impact test (LIPIT), detailed elsewhere (*9*, *16*). The test materials were nickel substrates of five different purities, 99.999%, 99.995%, 99.95%, 99.5% and 99%, with compositions given in Table 1; the solutes in these Ni samples are predominantly C, Fe, Cu, and Mn. These samples were tested at extreme strain rates by impacting them with hard alumina microspheres, 20 ± 1 µm in diameter, at two different temperatures, 20 and 155 °C. The velocities of our experiments ($v_i$ = 60-270 m/s) were chosen to be well below the regime where shock effects (such as jetting, hydrodynamic penetration, etc.) occur (~500 m/s (*17*)). For each material, the impact velocity, $v_i$, is plotted against the coefficient of restitution, CoR, or ratio of rebound to impact velocity ($v_r/v_i$), on a double logarithmic scale, Fig 1. In these experiments, the alumina impactors are effectively undeforming (elastic); higher CoR signifies that more energy is returned to the particle after impact and less is consumed as plasticity in the substrate. At the two different temperatures, for impacts at the same velocity, a faster rebound is a first indication of a higher strength metal.

For a fixed temperature as the impact velocity is increased, more energy is dissipated as plasticity, as seen through a drop in CoR with increasing impact velocity in Fig. 1. The parabolic scaling of the data in Fig. 1 is an expected form for plastic impacts, following the model of Wu et al. (*18*, *19*) for a contact with a single yield stress:

$$CoR = \frac{v_r}{v_i} = 0.78 \left( \frac{\sqrt{26} \cdot \sigma_y^{3/2}}{E^* v_i \sqrt{\rho}} \right)^{1/2} \quad (1)$$

where $\sigma_y$ is the dynamic yield strength, $E^*$ is the effective modulus between the impactor and substrate, and $\rho$ is the density of the impactor. The data in Fig. 1 can be fitted with Eq. (1) with a single parameter, $\sigma_y$. This value provides a measure of the average strength of the substrate over the duration of the impact. The fits are shown as solid lines in Fig. 1 for each set of data, along with the fitted dynamic yield strengths.

There are two interesting trends in Fig. 1. First, alloying increases strength as expected; comparing across the datasets, it is clear that as the purity level of the Ni decreases from left to right across the panels in the figure, the response curves rise and the corresponding fitted value of $\sigma_y$ increases. However, closer inspection reveals that this alloy strengthening effect is more potent at lower temperatures than at higher ones, meaning that the blue curves rise faster than the red ones in Fig. 1. In Fig. 1a-c a clear thermal strengthening, "hotter-is-stronger", trend is seen, where impacts rebound faster and the calculated $\sigma_y$ values are larger at the hotter test temperature. However, at the higher alloying levels (i.e., lower purity levels), the CoR curves overlap more clearly, and the strength values at the two temperatures become close. In Fig. 1d-e, a new trend is seen where the test carried out at higher temperatures now leads to thermal softening, "hotter-is-softer", again shown both as lower rebound velocities and lower calculated $\sigma_y$ values.

For each particle impact event, in addition to the in-situ observations of the particles in flight, a plastic indentation is left in the substrate. Each indentation is observed and measured using 3D laser scanning confocal microscopy. Across all test materials and temperatures, larger indentation volumes are a first indication of a softer material (i.e., more deformation). In Fig. 2, across the range of alloys, we see complementary trends in indentation volume to those observed earlier with $\sigma_y$. First, the addition of alloying elements leads to smaller indentations (hardening) across the series horizontally. Second, however, the higher purity samples have smaller indentation volumes with increasing temperature, and the lower purity samples have larger indentation volumes with increasing temperature. The two indentation volumes in Fig. 2b,e at the two test temperatures are approximately the same, consistent with the overlap seen in the corresponding CoR curves; this is roughly the point of the crossover from "hotter-is-stronger" to "hotter-is-softer" with the addition of alloying elements.

The impact indentation can be further used as the quantitative basis of the dynamic hardness, $H_d$, assessed as:

$$H_d = \frac{1/2 \cdot m_p \cdot (v_i^2 - v_r^2)}{V} \qquad (2)$$

where $V$ is the indentation volume, and $m_p$ is the mass of the impactor, measured using the density of the particle and its measured diameter. The dynamic hardness for each impact is plotted as a cumulative probability in Fig. 3, with the average strain rate across all impacts for each material and temperature in Table S1.

As with the CoR and $\sigma_y$ in Fig. 1, the hardness values in Fig. 3 show the same two major trends. First, alloying leads to higher hardness generally, as the cumulative distributions shift to the right with decreasing purity. Second, however, we see the same thermal crossover: Fig. 3 a-c show that the higher purity Ni is thermal hardening (hotter samples harder than cooler ones, red curve to the right of blue), while the lower purity Ni, Fig. 3d,e, is thermally softening, indicated by a flip in the trend of hardness data (blue curve to the right of red).

**Deformation Mechanisms**

The strength of a metal at strain rates above $10^6$ s$^{-1}$ can be approximated as the linear combination of the available strengthening mechanisms. In pure metals these include the thermally activated motion of dislocations at the intrinsic lattice strength (Peierls barriers, $\sigma_{th}$) and athermal strengthening from dislocation-grain boundary and dislocation-dislocation interactions ($\sigma_a$) (9, 15, 20–22); both of these classical mechanisms exhibit thermal softening. Additionally, there is a dislocation drag term that arises from dislocations interacting with phonons:

$$\sigma_d = \frac{B_d \dot{\varepsilon}}{\rho_m b^2} \quad (3)$$

where $b$ is the burgers vector, $B_d$ is the drag coefficient, $\dot{\varepsilon}$ is the strain rate, and $\rho_m$ is the mobile dislocation density. Notably, the drag term of Eq. (3) has antithermal behavior, i.e., stronger-is-hotter behavior, because $B_d$ increases linearly with temperature (21, 23–25). At the very high strain rates of our tests ~7×10$^6$ s$^{-1}$, the dislocation drag strengthening term of Eq. (3) can be dominant over the others in pure metals, giving rise to an overall increase in strength with increasing temperature (9). This was quantitatively established in our prior work on pure Cu, Au, and Ti with similar LIPIT experiments in (9), and the results in Figs. 1a and 3 extend the observation of thermal hardening to pure Ni as well; thermal strengthening is prima facie evidence that dislocation drag is the dominant mechanism in our experiments on pure Ni.

However, the addition of solute elements into a Ni matrix introduces a new strengthening mechanism that can both increase the overall magnitude of the strength and have a different scaling dependency with temperature. Solute atoms provide new "soft" pinning points for dislocations to interact with, which, like Peierls barriers, can be overcome by suitable thermal fluctuations. Solute strengthening is represented by a fourth additive contribution ($\sigma_s$) to the total strength $\sigma_y$:

$$\sigma_y = \sigma_{th} + \sigma_a + \sigma_d + \sigma_s \quad (4)$$

The role of solute strengthening has been addressed by many authors (26–28), and the present alloys are relatively dilute, and primarily incorporate interstitial elements (Table 1) such as C. While there are other solute elements in Ni, they contribute to substitutional strengthening at the negligible level of a few MPa, by comparison to interstitial elements which have a much larger effect due to their anisotropic lattice distortions. For such materials, the expression developed by Follansbee et al. (22) is appropriate for an illustrative analysis, and was calibrated by those authors to the specific case of Ni containing interstitial C:

$$\sigma_s = \left( \left[ 1 - \left( \frac{kT}{g_0 b^3 \mu_T} \ln\left(\frac{\dot{\varepsilon}_0}{\dot{\varepsilon}}\right) \right) \right]^{3/2} \sigma_c c \right)^n \quad (5)$$

Here $g_0$ is the normalized total free energy, $k$ is the Boltzmann constant, $b$ is the burgers vector, $\dot{\varepsilon}_0 = 10^{10}$ s$^{-1}$ is a reference strain rate, and $\mu_T$ is the temperature-dependent shear modulus. The exponent n = 1.4 speaks to the spacing of solutes in the path of dislocations. The critical stress, $\sigma_c$ = 20 GPa according to Follansbee et al. (*22*), and $c$ is the interstitial solute content.

Because this newly available strengthening term is positive and additive to the others, and yet is thermally controlled, it has two major effects: it strengthens, but it lowers the temperature dependence of strength (note that $d\sigma/dT$ is negative in Eq. (5)). As a result, pure samples, which are dominated at these high strain rates by dislocation drag ($\sigma_d$), have the highest anti-thermal hardening and the new $\sigma_s$ term is not critical to that temperature dependence. As solutes are added to the system, though, $\sigma_s$ increases its contribution, and the anti-thermal hardening slope is lessened. In the lowest purity samples, the slope has indeed flipped sign, reflecting a crossover to dominance of the $\sigma_s$ term, including its prevailing thermal activation characteristics.

The mechanistic cross-over described above can be seen in fuller form by evaluating Eq. (4), with the various strengthening terms populated with data for Ni as laid out above and in the supplemental materials. Figure 4 plots the experimentally assessed $\sigma_y$ values at the experimental strain rate of $10^7$ s$^{-1}$ as well as Eq. 4 using the full interstitial solute content for each alloy from Table 1. Both of the critical experimental trends are well captured by the model. First, solute elements cause strengthening at all rates, and the rise follows the expected scaling of Eq. (4) reasonably well. Second, the temperature dependence of strength changes as the solute drag term becomes dominant at higher solute concentrations. There is a critical concentration of solutes where, at extreme strain rates, the dominant mechanisms cross-over, and the strength of Ni becomes nominally independent of temperature. This crossover point lies at c ~ 0.4% for interstitials in Ni. This concentration also marks the point where Ni alloys transition from antithermal "hotter-is-stronger" to more typical thermally-activated "hotter-is-softer" deformation.

The transition we see in Fig. 4 is one that is expected to be common to many materials, because the competition amongst deformation mechanisms in Eq. (5) is common physics, although the details of the specific mechanisms and their functional forms may vary. For metals, the trend in Fig. 4 is broadly expected to prevail, and in supplemental Fig. S1 and S2 we demonstrate the generality of the present observations for two additional alloy families based on Ti and Au. In both cases, only the highest purity materials thermally hardened while the alloyed materials thermally softened. Because both of these materials exhibited the same behavior, one can infer that the "hotter-is-stronger" to "hotter-is-softer" transition seen is not specific to a single crystal structure (HCP Ti versus FCC Ni) or difference in surface oxides or structure (Au does not form a surface oxide under these conditions).

In summary, although models for high-rate deformation mechanisms in metals have existed for several decades, LIPIT accesses extreme rates, beyond $10^6$ s$^{-1}$, without strong shock effects dominating the material response. In addition to clean observations of phonon-drag dominated deformation, with their "hotter-is-stronger" behavior, here we illustrate the ability of LIPIT to map the range of dominance of that mechanism as it crosses over to other more conventional thermally-activated deformation mechanisms. Above strain rates of $10^5$ s$^{-1}$, highly pure metals increase strength and hardness with increasing temperature, while lower purity Ni thermally softens. This transition point from drag-controlled to solute-pinning-controlled mechanisms conforms to mechanistic strength models that account for both the strain rate and temperature effects of dislocation interactions with phonons as well as dislocation interactions with interstitials. There are many technological applications where such mechanistic transitions are of clear relevance, including hard particle erosion and sandblasting, high-speed subtractive manufacturing, cold dynamic gas spray manufacturing, and hypersonic transport. The present results suggest that when designing for, or protecting against, deformation at these rates, it is critical to consider all the mechanisms at play. Clearly, alloying can be used as a tool to precisely manipulate the temperature dependence of strength in unusual ways at these high rates, and there are compositions such as the c ~ 0.4% composition in Ni seen in Fig. 4 where the mechanical response is effectively temperature independent. We look forward to progress in the design of alloys for extreme conditions in light of the unusual deformation physics that prevail in microparticle impacts.

**Methods**

Nickel rods with nominal purity levels of 99.999%, 99.995%, 99.95%, 99.5% and 99% ThermoFisher, USA) were annealed at 500 °C for 7 hours before being ground flat and polished to a mirror finish using standard metallographic techniques. Alumina microparticles (Cospheric, Santa Barbara, CA) 20 ± 1 μm were used as indenters for impact experiments due to their high hardness compared to the nickel substrates (~15.7 vs. ~1.15 GPa at quasistatic rates)(*29*, *30*). 99.99% Ti (Alfa Aesar, USA) and a Grade 2 CP-Ti plate (Online Metals, USA) were annealed at 600 °C for 10 hours then ground and polished to a mirror finish. 99.999% and 99.95% Au rods (ThermoFisher) were rolled to an 85% reduction in thickness before being annealed at 125 °C for 3 hours. Note that the high purity samples of Ti and Au are the same materials used in our prior work (*9*), and we reproduce the data for those samples here, alongside new data for the alloyed samples of Ti and Au.

Microparticle impact experiments were carried out using optically driven microballistics, detailed elsewhere (*16*, *31*), over impact velocities ranging from 30–270 m/s. A monolayer of microparticles was finely dispersed on a glass-sandwich launching pad consisting of a 200 μm glass slide, 60 nm chromium layer, UV curable glue, and a second 200 μm glass slide. A single particle was preselected prior to launch based on its size, shape, and surface morphology. To extend these experiments to elevated temperatures, a resistive heating stage was put in place to heat the substrate, particles and surrounding vapor to a uniform temperature, checked with a series of thermocouples. The stand–off distance between the particles and substrate was 750 μm. The particle was placed behind a single laser pulse (532 nm, 10 ns, Nd:YAG) causing an ablation of the chromium and the microparticle to be accelerated towards the test target. The flight, impact, and rebound was captured with a high–speed camera (SIMX-16, Specialised Imaging), capable of producing a 16–frame image sequence. The impact and rebound velocities of the particle were measured between frames and the net strain rate of each impact was calculated to as the impact velocity divided by the particle diameter ($v_i/d$). This strain rate represents the net strain rate over the duration of the impact and is taken as the average strain rate over the entire impact event. Post mortem analysis of the indentations was done using a laser scanning confocal microscope (VK-X200, Keyance) to measure indentation volumes.

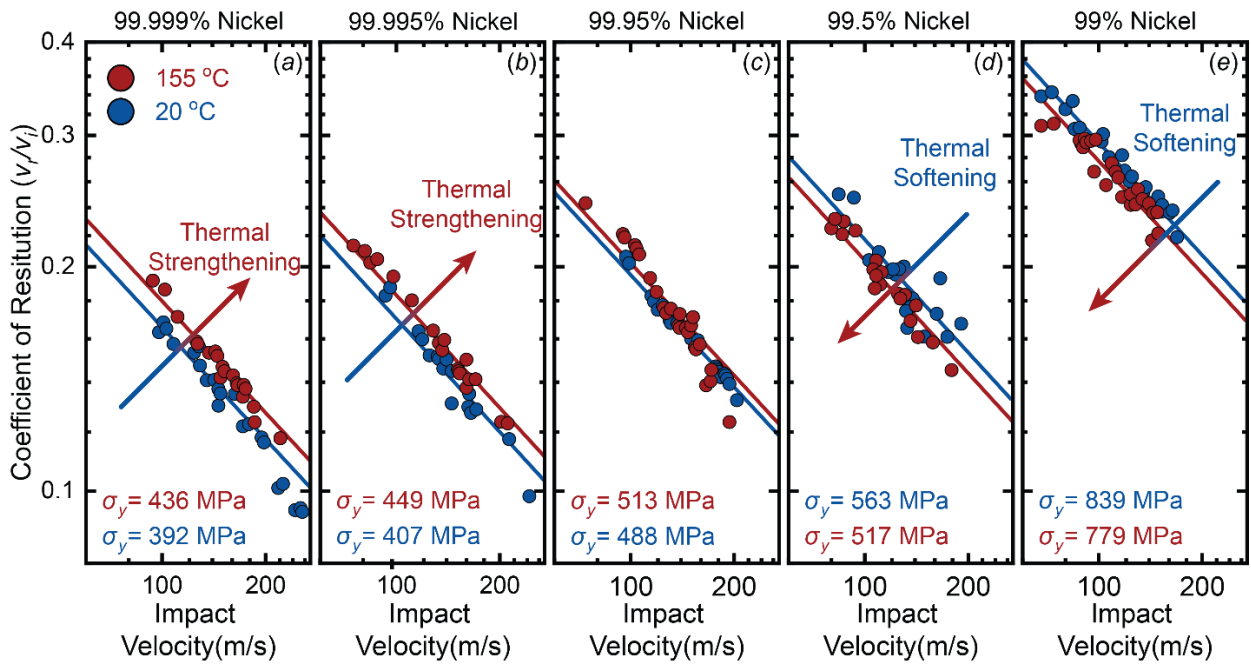

Figure 1: The impact velocity verses coefficient of restitution is plotted on a double logarithmic scale for **(a)** 99.999% Ni **(b)** 99.995% Ni **(c)** 99.95% Ni **(d)** 99.5% Ni and **(e)** 99% Ni at 20 °C and 155 °C. There is a clear change from thermal hardening to thermal softening when increasing the solute content in Ni with 99.95% Ni serving as the approximate transition point. Solid lines indicate the parabolic scaling of the impact and rebound behavior as proposed by Wu et al. This scaling was used to calculate the dynamic yield strength $\sigma_y$ for each material at each temperature.

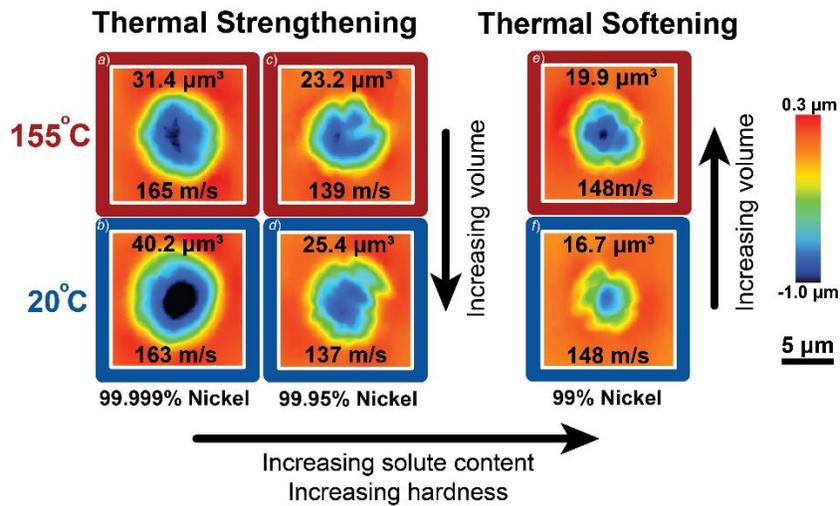

Figure 2: Laser scanning confocal micrographs are shown for **(a,b)** 99.999% Ni, **(c,d)** 99.95% Ni, and **(e,f)** 99% Ni with inserts showing the measured crater volume and impact velocities. The crater volume increases from **(a)** to **(b)** indicating the expected thermal hardening in pure metals. The crater volume decreses from **(e)** to **(f)** indicating classic thermal softening in the lowest purity material. The crater volumes in **(c)** to **(d)** are approximately equal corresponding to the transition point from thermal hardening to thermal softening. What is more, as the purity level decreases, the crater volume decreases, another expected form of plasticity where alloy elements increase the hardness of a metal.

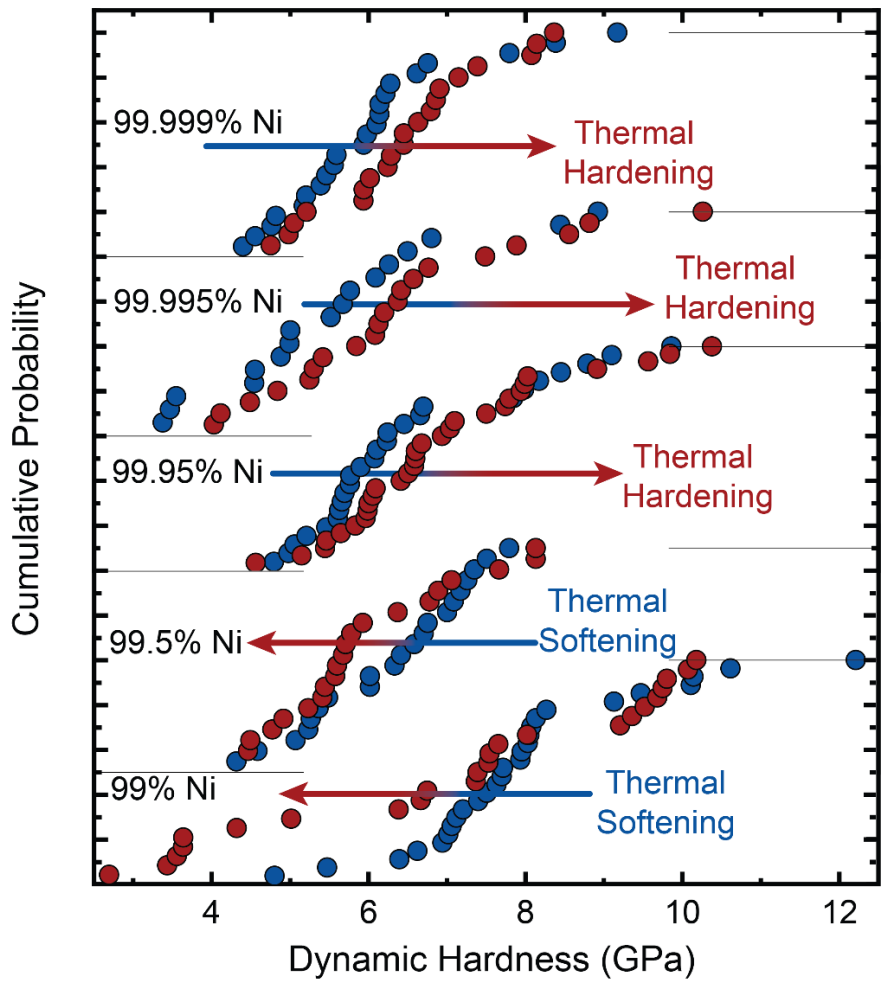

Figure 3: The dynamic hardness for each impact experiment was calculated and plotted as a cumulative probability for each test material and temperature. As with the dynamic yield strength and measured crater volumes, the highest purity samples exhibited thermal hardening and the lowest purity samples exhibited thermal softening. Additionally, the average hardness at 20 °C is increasing with increasing solute content, the expected form of plasticity.

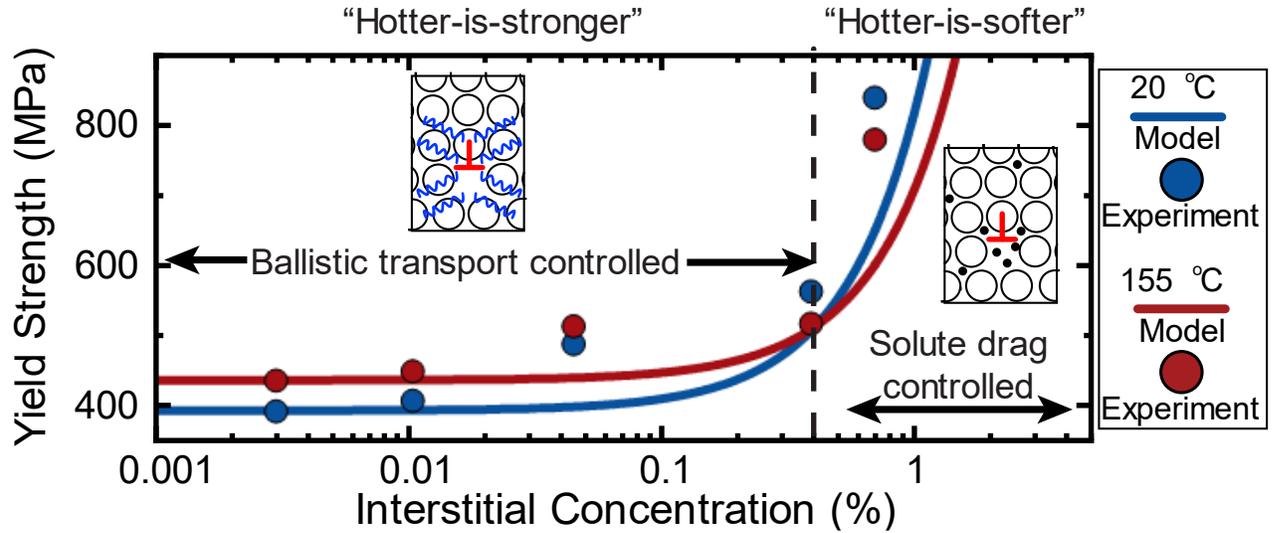

Figure 4: The experimentally assessed dynamic yield strength for each test material is plotted as a function of interstitial concentration, filled circles. As the interstitial concentration increases, for each temperature, the yield strength increases. Solid lines show the total yield strength predicted from the model, including the solute strengthening term from Eq. 5. As the interstitial concentration increases from 0.001%, the materials exhibit thermal hardening then cross over to thermal softening at an interstitial concentration of ~0.4%. This cross over from "hotter-is-stronger" to "hotter-is-softer" arises from a change in the dislocation kinetics from ballistic transport of dislocations to solute drag controlled, which is thermally activated.

| Name | Ni | C | Mn | Cu | Fe | Si | S | B | Co | O | Total Interstitial |
|---|---|---|---|---|---|---|---|---|---|---|---|
| **99.999** | 99.997 | 0.003 | $1.07 \times 10^{-7}$ | $1.29 \times 10^{-5}$ | $2.5 \times 10^{-4}$ | $6.3 \times 10^{-6}$ | - | - | - | - | 0.003 |
| **99.995** | 99.986 | 0.01 | $2.7 \times 10^{-6}$ | $7.4 \times 10^{-6}$ | 0.001 | $8.3 \times 10^{-6}$ | 0.0004 | - | - | - | 0.0104 |
| **99.95** | 99.95 | 0.037 | - | - | - | - | 0.003 | 0.005 | 0.001 | - | 0.045 |
| **99.5** | 99.24 | 0.39 | 0.17 | 0.018 | 0.01 | 0.17 | $3.6 \times 10^{-4}$ | - | - | - | 0.39 |
| **99*** | 98.54 | 0.24 | 0.14 | 0.027 | 0.010 | 0.35 | 0.0018 | 0.016 | - | 0.44 | 0.698 |

Table 1: The composition of each alloy given in atomic percent. The substitutional elements (Mn, Cu, Fe, Si, and Co) have very little impact on alloy strength, while the interstitial elements (C, B, O, and S) have strong interactions with dislocations; their total composition in the final column is used in the model.

# Supplementary Materials: *At extreme strain rates, pure metals thermally harden while alloys thermally soften*

## S1. Thermal hardening to softening cross over in other metals

In addition to Ni, the transition from thermal hardening to thermal softening with increasing solute content is seen in other metals – here Ti and Au. Ti was chosen for its hexagonally close packed crystal structure (as a complement to FCC Ni), while Au was chosen because it does not form a native oxide layer on its surface to rule out any oxide effects that might have occurred due to heating the materials for impact tests.

Two test materials of Ti were chosen, high purity 99.99% Ti and commercially pure (CP-Ti), Table S2. Data for 99.99% Ti was previously published in our recent work in Ref. (*9*) and is reproduced here; data for CP-Ti are new to this work. Coefficient of restitution plots for alumina impactors on each test material are shown in Fig. S1 a,b. Fitting the impact and rebound data to the Wu et al. plasticity model (*19*) at 20 ºC, 99.99% Ti has a lower dynamic strength than CP-Ti: 776 MPa compared to 1029 MPa. However, the 99.99% Ti sample thermally strengthens by 15% when heated to 200 ºC while the CP-Ti sample thermally softens by 10% under the same impact conditions. This same trend is seen in Fig. S1c in the dynamic hardness cumulative probabilities. As with the strength, the hardness for 99.99% Ti is lower than the hardness for CP-Ti and exhibits thermal hardening, while the hardness for CP-Ti shows a clear thermal softening trend.

Two different gold specimens were chosen as test materials, 99.999% Au and 99.95% Au, Table S3. Data for the 99.999% Au material was again taken from our recent work Ref. (*9*), and reproduced here; data for 99.95% Au is new to this work. In Figure S2a,b, the higher purity Au sample exhibited thermal strengthening while the lower purity sample thermally softened when using the Wu et al. plasticity model for dynamic strength (*19*). This trend is again reaffirmed using the dynamic hardness measurements and crater volumes. Because the temperature difference is smaller for these samples, only ~80 ºC, there is less of a shift in the hardness and strength when compared to the other materials.

## S2. Dynamic Strength Model

The strength of a metal can be estimated as the linear combination of its individual available strengthening mechanisms. In pure metals at extreme strain rate deformations, above $10^5$ s$^{-1}$, these are the thermal strengthening ($\sigma_{th}$), athermal strengthening ($\sigma_a$), phonon drag strengthening ($\sigma_d$), and solute strengthening ($\sigma_s$). While the latter two terms are full described in the main text, the former two terms can be evaluated as follows.

### S2.1 Thermal Strengthening

The thermal strengthening mechanism comes from dislocation interactions with thermally activated barriers in the crystal lattice: i.e. internal lattice frictions, Peierls barriers, and are fully overcomable by thermal fluctuations alone – leading to a decrease in strength with increasing temperature.

$$\sigma_{th} = \left[1 - \left(\frac{kT}{g_0 \mu_T b^3} \ln\left(\frac{\dot{\varepsilon}_0}{\dot{\varepsilon}}\right)\right)^{1/q}\right]^{1/p} \sigma_0 \quad \text{(S1)}$$

$$\mu_T = \mu_o - \frac{s}{e^{\left(\frac{T_o}{T}\right)} - 1} \quad \text{(S2)}$$

In Eq S1, $\sigma_o$ is the stress needed to overcome the short-range barriers at 0 K and $q = 2/3$ and $p = 2$ are fitting parameters based on the energy shape of the barriers. The temperature dependent shear modulus, Eq. S2, can be calculated using the model by Varshni (*32*), where $T_o$ and $s$ are constants equal to 258 K and 8.839 GPa and $\mu_o$ is the shear modulus at 0 K, 84.52 GPa, for pure Ni.

## S2.2 Athermal Strengthening

Dislocations also interact with other dislocations and grain boundaries, providing athermal strengthening contributions. The magnitude increase in strength of these interactions is much greater than with short-range barriers and, importantly, cannot be overcome by thermal fluctuations alone.

$$\sigma_a = \frac{\alpha_G \mu_T \sqrt{b}}{\sqrt{D}} + \alpha_{disl} \mu_T b \sqrt{\rho} \quad \text{(S3)}$$

The interaction of dislocations with grain boundaries is modeled here using a simple Hall-Petch like relation, the first term in Eq. S3, where $D$ is the grain size after impact, estimated to be on the order of several hundred nanometers from the work of Ref. (*33*), and $\alpha_G$ is an interaction parameter between dislocations and grain boundaries taken to be 0.15. The second term of Eq. S3 captures the strengthening from dislocation-forest and dislocation-dislocation interactions through the mobile dislocation density, $\rho$ (~$10^{13}$ m$^{-2}$), and the dislocation-dislocation interaction parameter $\alpha_{disl} = 0.5$. While these barriers cannot be overcome by thermal fluctuations alone, there is a small decrease in strength in each term with temperature due to a decrease in the temperature dependent shear modulus.

| | Temperature | Average Strain Rate (× $10^6$) | Standard Deviation (× $10^6$) |
|---|---|---|---|
| 99.999% Ni | 20 °C | 7.77 | 2.14 |
| | 155 °C | 7.61 | 1.56 |
| 99.995% Ni | 20 °C | 7.29 | 1.85 |
| | 155 °C | 6.73 | 1.83 |
| 99.95% Ni | 20 °C | 7.69 | 1.61 |
| | 155 °C | 6.86 | 1.36 |
| 99.5% Ni | 20 °C | 6.57 | 1.41 |
| | 155 °C | 5.92 | 1.30 |
| 99% Ni | 20 °C | 5.96 | 1.47 |
| | 155 °C | 5.68 | 1.21 |

Table S1: The average strain rate and standard deviation for impacts onto each material.

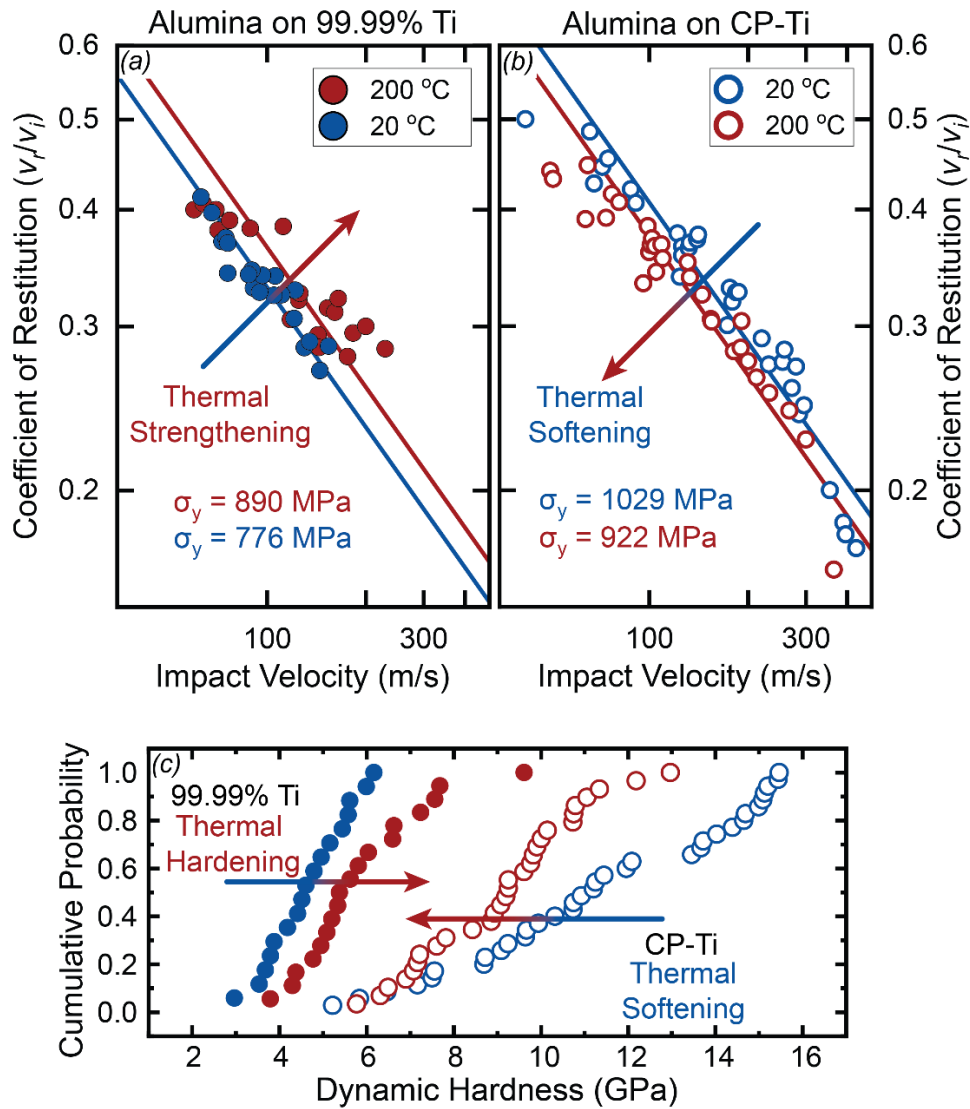

Figure S1: The impact velocity verses coefficient of restitution is plotted on a double logarithmic scale for **(a)** 99.99% Ti **(b)** CP-Ti at 20 °C and 200 °C. Solid lines indicate the parabolic scaling of the impact and rebound behavior as proposed by Wu et al. This scaling was used to calculate the dynamic yield strength $\sigma_y$ for each material at each temperature. **(c)** The dynamic hardness for each impact experiment was calculated and plotted as a cumulative probability for each test material and temperature. As with the dynamic yield strength, the highest purity samples exhibited thermal hardening and the lowest purity samples exhibited thermal softening.

| | Ti | N | C | H | Fe | O |
|---|---|---|---|---|---|---|
| **High Purity Ti** | 99.82 | $6.8 \times 10^{-4}$ | 0.029 | 0.016 | $7.6 \times 10^{-4}$ | 0.132 |
| **CP-Ti Grade 2** | 99.3 | 0.03 | 0.1 | 0.015 | 0.3 | 0.25 |

Table S2: The atomic percentages of the high purity Ti rod and Grade 2 commercially pure Ti plate are given below.

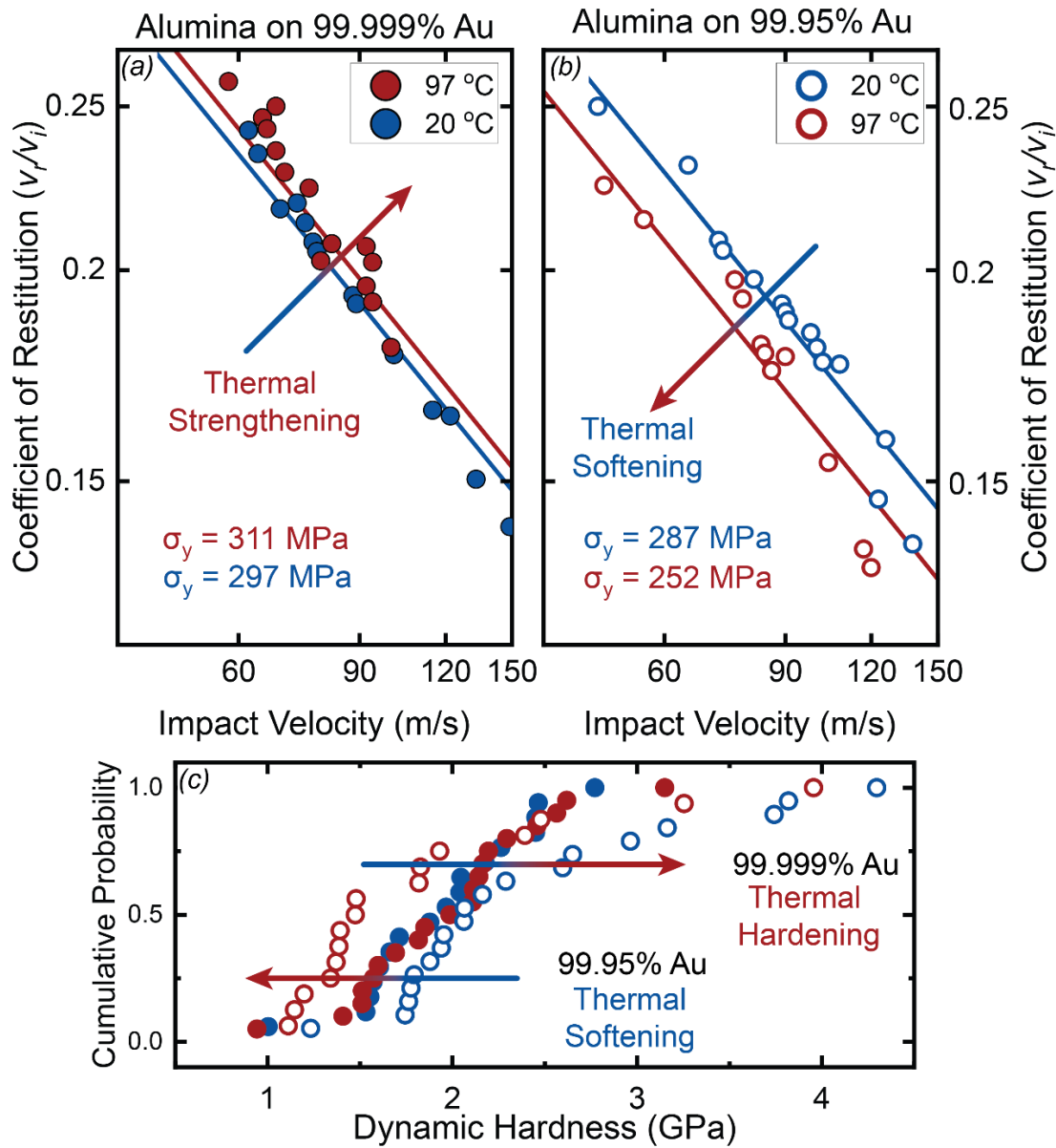

Figure S2: The impact velocity verses coefficient of restitution is plotted on a double logarithmic scale for **(a)** 99.999% Au **(b)** 99.95% Au at 20 °C and 97 °C. Solid lines indicate the parabolic scaling of the impact and rebound behavior as proposed by Wu et al. This scaling was used to calculate the dynamic yield strength $\sigma_y$ for each material at each temperature. **(c)** The dynamic hardness for each impact experiment was calculated and plotted as a cumulative probability for each test material and temperature. As with the dynamic strength, the hardness transitions from thermal hardening to thermal softening in the less pure sample.

| | Au | Ag | Ca | Cu | Ti | Pt |
|---|---|---|---|---|---|---|
| **99.99% Au** | 99.999 | $7.1 \times 10^{-4}$ | $3.9 \times 10^{-4}$ | $6.2 \times 10^{-5}$ | $4.5 \times 10^{-4}$ | $4.0 \times 10^{-5}$ |
| **99.95% Au** | 99.97 | $9.8 \times 10^{-3}$ | $2.5 \times 10^{-3}$ | 0.015 | - | 0.001 |

Table S3: The composition of each Au test material is given below in atomic percent.